\documentclass[showpacs,preprintnumbers,preprint,aps,amsmath,amssymb]{revtex4-1}
\pdfoutput=1
\usepackage{color}
\usepackage{graphics}
\usepackage{amsmath,amssymb,epsfig}
\newcommand{\mathsym}[1]{{}}

\newcommand{\ba}{\begin{array}}
\newcommand{\ea}{\end{array}}
\newcommand{\be}{\begin{equation}}
\newcommand{\ee}{\end{equation}}
\newcommand{\beqa}{\begin{eqnarray}}
\newcommand{\eeqa}{\end{eqnarray}}
\def\321{$SU(3)\times SU(2)\times U(1)$}
\def\b126{$\overline{126}$}
\def\mt{$\mu$-$\tau$ }
\def\mnuf{${\cal M}_{\nu f }$~}

\newcommand{\dms} {\Delta m^2_{sol}}
\newcommand{\Dma} {\Delta m^2_{atm}}

\begin{document}
\vspace*{1cm}
\title{Minimal extension of tri-bimaximal mixing and generalized $Z_2 \times Z_2$ symmetries}
\bigskip
\author{Shivani Gupta\footnote{shivani@prl.res.in}, Anjan S. Joshipura\footnote{anjan@prl.res.in}
and Ketan M. Patel\footnote{kmpatel@prl.res.in}}
\affiliation{Physical Research Laboratory, Navarangpura, Ahmedabad-380 009,
India. \vskip 1.0truecm}

\begin{abstract}
We discuss consequences of combining the effective  $Z_2\times Z_2$ symmetry of the tri-bimaximal
neutrino mass matrix with the CP symmetry. Imposition of such generalized  $Z_2\times Z_2$
symmetries leads to predictive neutrino mass matrices determined in terms of only four parameters
and leads to non-zero $\theta_{13}$ and maximal atmospheric mixing angle and CP violating phase. It
is shown that an effective generalized $Z_2\times Z_2$  symmetry of the mass matrix can arise from
the $A_4$ symmetry with specific vacuum alignment. The neutrino mass matrix in the considered model
has only three real parameters and leads to determination of the absolute neutrino mass scale as a
function of the reactor angle $\theta_{13}$.                                                        
                                            
\end{abstract}

\maketitle

\section{Introduction}
Recent $\nu_e-\nu_\mu$ oscillation observations by T2K \cite{t2k} and MINOS \cite{minos} and double CHOOZ \cite{dc} have led to
a search of alternatives to the Tri-bimaximal (TBM) leptonic mixing \cite{hs} pattern among
neutrinos. 
\be \label{utbm}
U_{TBM}=\left( \ba{ccc} \sqrt{\frac{2}{3}}&\frac{1}{\sqrt{3}}&0\\
		  -\frac{1}{\sqrt{6}}&\frac{1}{\sqrt{3}}&-\frac{1}{\sqrt{2}}\\
                  -\frac{1}{\sqrt{6}}&\frac{1}{\sqrt{3}}&\frac{1}{\sqrt{2}} \\ \ea \right) \ee 
The above pattern corresponds to $\sin^2\theta_{12}=1/3$, $ \sin^2\theta_{23}=1/2$ and
$\sin^2\theta_{13}=0$ for the three mixing angles involved in neutrino oscillations. It is
theoretically well founded and can be obtained using flavour symmetries in the leptonic sector, see
\cite{af} for a review and original references. While the predicted values of the $\theta_{12}$ and
$\theta_{23}$ in TBM agree nearly within 1$\sigma$ of the latest global analysis \cite{fogli,valle}
of the neutrino oscillation data, prediction $\theta_{13}=0$ is at variance with T2K \cite{t2k}
(MINOS \cite{minos}) results by $2.5\sigma (1.6\sigma)$ and with the global analysis
\cite{fogli,valle} by about 3$\sigma$. This suggests that one should look either for perturbations
to TBM affecting mainly $\theta_{13}$ or try to look for alternative flavour symmetries which imply
non-zero $\theta_{13}$. Recently, several attempts \cite{modifiedTBM,hezee,models} have been made in
this directions. Some of these works \cite{modifiedTBM,hezee} discuss the possible schemes of
perturbations to TBM while some \cite{models} provide the models also. The minimal scheme
would be the one in which $\theta_{13}$ gets generated but $\theta_{23}$ and $\theta_{12}$ remain
close to their predictions in the TBM scheme. We show that this can be achieved by generalizing the
$Z_2\times Z_2$ symmetry of the TBM mass matrix and identify appropriate flavour symmetry which can
lead to the
modified pattern.

The paper is organized as follows. In the next section, we discuss the generalized $Z_2\times Z_2$
symmetry of neutrino mass matrix which minimally modifies the TBM mixing pattern and leads to
nonzero $\theta_{13}$. In Section \ref{sec:model}, we present possible modifications in well know
$A_4$ model which lead to the neutrino mass matrix invariant under the proposed symmetry and discuss
its phenomenology. Finally, we summarize in Section \ref{sec:summary}.

\section{Generalized $Z_2\times Z_2$ symmetry and lepton mixing angles}
\label{sec:generalizedsymm}

A well-known property of the TBM pattern is the presence of a specific $Z_2\times Z_2$ symmetry
\cite{lam} enjoyed by the corresponding  neutrino mass matrix \mnuf in the flavour basis. This symmetry is defined  in
general by the operators $S_{i}~,i=1,2,3$ :
\be \label{si}
(S_i)_{jk}=\delta_{jk}-2 U_{ji}U_{ki}^*~,\ee
where $U$ is the matrix diagonalizing \mnuf. Each $S_i$ defines a $Z_2$ group and satisfies  
\be \label{relation}
S_iS_j=-S_k~~,~~~i\not=j\not=k~\ee
The $S_i$ also leave the neutrino mass matrix invariant
\be \label{invariance} 
S_i^T{\cal M}_{\nu f} S_i={\cal M}_{\nu f}~.\ee
as can be verified from the Eq. (\ref{si}) and the property
${\cal M}_{\nu f}=U^*D_{\nu}U^{\dagger}$, $D_{\nu}$ being the diagonal neutrino mass matrix.

The explicit forms for $S_2$ and $S_3$ in the TBM case are given by:
\be \label{s2s3} \ba{cc}
S_2=\dfrac{1}{3}\left( \ba{ccc} 1&-2&-2\\
		  -2&1&-2\\
                  -2&-2&1 \\ \ea \right) ~~~{\rm and}~~~
&
S_3=\left( \ba{ccc} 1&0&0\\
		  0&0&1\\
                  0&1&0 \\ \ea \right).\ea \ee
$S_2$ and $S_3$ respectively are determined by the second and the third column of the TBM mixing
matrix (\ref{utbm})
using Eq. (\ref{si}). The element $S_1$ can be obtained using relation (\ref{relation}). In
particular, $S_3$ corresponds to the well-known $\mu$-$\tau$ symmetry which is responsible for two
of the three predictions namely $\theta_{13}=0,\theta_{23}=\frac{\pi}{4}$ of the TBM pattern.

A desirable replacement of the $\mu$-$\tau$ symmetry would be the one which retains maximality of 
$\theta_{23}$ but allows a non-zero $\theta_{13}$. Such a symmetry is already known \cite{grimus1}
and is obtained by combining the $\mu$-$\tau$ symmetry with the CP transformation. The neutrino mass
matrix gets transformed to its complex conjugate under the action of the generalized $\mu$-$\tau$:
\be \label{gens3} 
S_3^T{\cal M}_{\nu f} S_3={\cal M}_{\nu f}^*~.\ee
A neutrino mass matrix with this property leads to two predictions \cite{grimus1}:
\beqa \label{z2predictions}
\sin^2 \theta_{23} &=& \frac{1}{2}, \\
\sin \theta_{13} \cos \delta &=& 0.
\eeqa
One needs a non-zero $\theta_{13}$ in which case, the above equation leads to  a prediction $\delta=\frac{\pi}{2}$ for the CP violating 
Dirac phase. Eq. (\ref{gens3}) does not put
any restrictions on the solar angle. In order to do this, we would like to combine the
generalized $\mu$-$\tau$ symmetry with the ``magic symmetry'' corresponding to invariance under
$S_2$ and define a generalized $Z_2\times Z_2$ symmetry. This can be done in two independent ways. 

{\flushleft {\bf Case I:}}~~ Let us first assume that the neutrino mass matrix in flavour basis
simultaneously satisfies
\be \label{case1}
S_{1,3}^T {\cal M}_{\nu f } S_{1,3} = {\cal M}_{\nu f }^*.
\ee
Both these conditions together imply  that
\be \label{s2}
S_{2}^T {\cal M}_{\nu f } S_{2} = {\cal M}_{\nu f }.
\ee
The above condition fixes the second column of the PMNS matrix $U$ to be $1/\sqrt{3}(1,1,1)^T$. This
form of $U$ is studied before and known as tri-maximal mixing pattern
\cite{trimaximal1,trimaximal2}. When compared with the standard form \cite{pdg}, it leads to the
relation
\be \label{solar1}
|\sin\theta_{12} \cos\theta_{13}|=\frac{1}{\sqrt{3}} ~~\Longrightarrow~~ \sin^2
\theta_{12}=\frac{1}{3}(1+\tan^2 \theta_{13}).
\ee
which provides the lower limit $\sin^2\theta_{12}\geq1/3$. The neutrino mass matrix in the flavour
basis \mnuf that satisfies (\ref{case1}) can be written as
\be \label{mnuf1}
{\cal M}_{\nu f}=
\left(
\ba{ccc}
y+z-x&x+i x'& x-i x'\\
x+i x'&y-i x'& z\\
x-i x'&z&y+i x'\\
\ea
\right),~
\ee
where all the parameters are real. Note that Re(${\cal M}_{\nu f}$) is invariant under
(\ref{invariance}) and so it is in the TBM form while Im(${\cal M}_{\nu f}$) follows
the condition
$$ S_{1,3}^T~{\rm Im({\cal M}_{\nu f})}~S_{1,3} = -{\rm Im({\cal M}_{\nu f})} $$\\
The neutrino mass matrix in Eq. (\ref{mnuf1}) can be diagonalized by the matrix 
\be \label{pmns1}
U^I=U_{TBM} P R_{13}(\theta)~,\ee
where $P={\rm diag.} (1,1,i)$ and $R_{13} (\theta)$ denotes a rotation by an angle $\theta$ in the $1-3$ plane. 

{\flushleft {\bf Case II:}}~~ The second possibility is 
\be \label{case2}
S_{2,3}^T {\cal M}_{\nu f } S_{2,3} = {\cal M}_{\nu f }^*~,
\ee
which leads to
\be \label{s1}
S_{1}^T {\cal M}_{\nu f } S_{1} = {\cal M}_{\nu f }.
\ee
This fixes the first column of $U$ to be $1/\sqrt{6}(2,-1,-1)^T$ which implies 
\be \label{solar2}
|\cos\theta_{12} \cos\theta_{13}|=\sqrt{\frac{2}{3}} ~~\Longrightarrow~~ \sin^2
\theta_{12}=\frac{1}{3}(1-2\tan^2 \theta_{13}).
\ee
In contrast to the previous case, this provides an upper bound on the solar angle
$\sin^2\theta_{12}\leq1/3$. The neutrino mass matrix in the flavour basis \mnuf in this case can be
written as
\be \label{mnuf2}
{\cal M}_{\nu f}=
\left(
\ba{ccc}
y+z-x&x+i x'& x-i x'\\
x+i x'&y+2i x'& z\\
x-i x'&z&y-2i x'\\
\ea
\right)~.
\ee
The above ${\cal M}_{\nu f}$ can be diagonalized by the matrix 
\be \label{pmns2}
U^{II}=U_{TBM} P R_{23}(\theta)~,\ee
where $R_{23} (\theta)$ denotes a rotation by an angle $\theta$ in the $2-3$ plane. 
The third possibility is to consider $ S_{1,2}^T {\cal M}_{\nu f } S_{1,2} = {\cal M}_{\nu f}^* $
and
this results into the \mt symmetric \mnuf which leads to $\theta_{13}=0$, so it is not the case of
our interest. Both the above scenarios predict small deviations in $\sin^2\theta_{12}$ from its
tri-bimaximal value, but in opposite directions. While both of them are consistent with the present
3$\sigma$ ranges of $\theta_{12}$ and $\theta_{13}$ obtained from the global fits to the recent
neutrino oscillation data \cite{valle}, prediction (\ref{solar2}) is more favored if only
1$\sigma$ is considered. Note that both these scenarios lead to the trivial Majorana phases (0
or $\pi$) and do not restrict the masses of neutrinos. 

The mass matrices in Eqs. (\ref{mnuf1},\ref{mnuf2}) based on the generalized $Z_2\times Z_2$
symmetry are different and more predictive 
compared to most other proposed  modifications of the TBM structure \cite{modifiedTBM,models,hezee}.
Let us emphasize the main differences:
\begin{itemize}
\item Eqs. (\ref{mnuf1},\ref{mnuf2}) contain four real parameters and hence lead to five predictions
among nine
observables. These are two trivial Majorana phases, a Dirac phase $\delta=\pm \pi/2$, an
atmospheric mixing angle $\theta_{23}=\pi/4$ and the solar mixing angle predicted by
Eq. (\ref{solar1}) or (\ref{solar2}).
\item Grimus and Lavoura \cite{trimaximal1} and He and Zee \cite{hezee} proposed a mixing matrix
similar to Eq. (\ref{pmns1}).
The  differences being the absence of $P$, presence of the Majorana phase matrix and the  replacement of  $R_{13}$  by a unitary 
transformation in the $1-3$ plane with an undetermined Dirac CP phase $\delta$. In the process, $\delta$ and Majorana phases become 
unpredictable and $\theta_{23}$ deviates from the TBM value by a term of ${\cal O}(\theta_{13})$.
 \item Likewise, Ma in his classic paper \cite{maclassic} considered  a modification to TBM
analogous to Eq. (\ref{pmns2}).
Here also $R_{23}$ gets replaced by a unitary transformation in the $2-3$ plane with an undermined phase resulting in less predictivity than the present case.
\item  A special case of Eq. (\ref{mnuf1}) was considered by Grimus and Lavoura \cite{trimaximal2}.
This corresponds to choosing
$$x' = -\frac{1}{\sqrt{3}} (z-x).$$ 
As a result, ${\cal M}_{\nu f}$ contains  only three parameters and allows determination of the absolute neutrino mass scale in addition to the  five predictions mentioned above. 
It is also shown in \cite{trimaximal2} that
such a mass matrix can arise in a model based on the $\Delta_{27}$ group. 
So far we have not appealed to any flavour symmetry at the Lagrangian level and considered 
only the effective $Z_2\times Z_2$ symmetry of the neutrino mass matrix. We now propose to realize this 
effective symmetry from an underlying flavour symmetry. In the process, we find that the use of 
flavour symmetry also leads us to a three parameter neutrino mass matrix as  in the case 
proposed by Grimus and Lavoura \cite{trimaximal2}.
\end{itemize}
\section{Model and  Phenomenology}
\label{sec:model}
We use the flavour symmetry $A_4$. Several versions of this symmetry  are proposed \cite{af}  to obtain a neutrino mass matrix 
which exhibits the TBM mixing. Here we show that a simple modification of the existing $A_4$ schemes
can lead to more predictive mass matrix given in Eq. (\ref{mnuf2}).  For definiteness, we
concentrate on 
a specific $A_4$ model of  He, Keum and Volkas
\cite{heA4}. We propose two possible schemes one based on the type-I seesaw and the other using a combination 
of both the type-I and type-II seesaw mechanisms.

Let us first outline the basic features of $A_4$ model proposed in \cite{heA4}. Though it was
proposed to explain both quark and lepton mixing patterns, we here discuss only the lepton sector
of it. The matter and Higgs field content of the model with their assignments under the SM gauge
group $G_{SM}\equiv SU(3)_c\times SU(2)_L \times U(1)_Y$ and $A_4$ group are given in
Table \ref{tab:1}.\\
\begin{table} [h]
\begin{math}
 \begin{array}{|c||c|c|c|c|c||c|c|c|}
\hline
 & l_L & e_R & \mu_R & \tau_R & \nu_R & \Phi & \phi & \chi \\
\hline
G_{SM} & (1,2,-1) & (1,1,-2) & (1,1,-2) & (1,1,-2) & (1,1,0) & (1,2,-1) & (1,2,-1)&
(1,1,0) \\
\hline
A_4 & {\bf3} & {\bf1} & {\bf1'} & {\bf1''} & {\bf3} & {\bf3} & {\bf1} & {\bf3} \\
\hline
\end{array}
\end{math}
\caption{Various fields and their representations under $G_{SM} \times A_4$.}
\label{tab:1}
\end{table}

The renormalizable $G_{SM} \times A_4$ Yukawa interactions of the model can be written as
\beqa \label{yukawa}
-{\cal L}_{Y}&=& y_e (\overline{l}_L {\tilde \Phi})_{\bf1} e_R+y_{\mu} (\overline{l}_L {\tilde
\Phi})_{\bf1''} {\mu}_R + y_{\tau} (\overline{l}_L {\tilde \Phi})_{\bf1'} {\tau}_R\nonumber \\
&+&y_D  (\overline{l}_L {\nu}_R)_{\bf1} \phi\nonumber \\
&+&\frac{1}{2} M \overline{\nu}_R \nu_R^c + \frac{1}{2} B' (\overline{\nu}_R \nu_R^c)_{\bf 3} \chi +
h.c., \eeqa
where ${\tilde \Phi}\equiv i \tau_2 \Phi^*$ and $(..)_R$ denotes $R$-dimensional representation of
$A_4$. Note that in \cite{heA4}, an additional $U(1)_X$ symmetry is also imposed so that an unwanted
$G_{SM} \times A_4$ invariant term $\overline{l}_L {\nu}_R \Phi$ can be forbidden when it is assumed
that $l_L,e_R,\mu_R,\tau_R$ and $\phi$ carry  $X=1$ and other fields are chargeless under $U(1)_X$.
Specific choice of the $A_4$ vacuum  $\langle \Phi \rangle = \upsilon (1,~1,~1)^T$  leads to the charged lepton mass matrix:
\be \label{clmass}
M_l=\sqrt{3} \upsilon~ U(\omega)~{\rm Diag.}(y_e,~y_{\mu},~y_{\tau})~,
\ee
where 
\be \label{Uomega}
U(\omega)=\frac{1}{\sqrt{3}}
\left(
\ba{ccc}
1&1&1\\
1&\omega^2& \omega\\
1&\omega&\omega^2\\
\ea
\right)
\ee
and $\omega=e^{ 2i\pi/3}$ is a cube root of unity. The Dirac neutrino mass matrix is
proportional to the identity matrix
\be \label{diracmass}
M_D=y_D \upsilon_\phi {\bf I}, \ee
where $\upsilon_{\phi} = \langle \phi \rangle$. Further, assuming that the field $\chi$ develops
a vacuum expectation value (vev)  in the direction $\langle \chi \rangle = \upsilon_{\chi} (1,~0,~0)^T$, the heavy neutrino mass
matrix can be written as
\be \label{MR}
M_R=\left(
\ba{ccc}
A&0&0\\
0&A & B\\
0&B & A\\
\ea
\right),
\ee
where $B=B' \upsilon_{\chi}$. After the seesaw, Eq. (\ref{diracmass}) and (\ref{MR}) lead to the
light neutrino mass matrix
\be \label{mnutbm}
{\cal M}_{\nu}= -M_D M_R^{-1} M_D^T =  \left(
\ba{ccc}
\frac{(a^2-b^2)}{a}&0&0\\
0&a & b\\
0&b & a\\
\ea
\right),
\ee
where $a=-\dfrac{y_D^2 \upsilon_{\phi}^2}{A^2 - B^2} A$ and $b=\dfrac{y_D^2 \upsilon_{\phi}^2}{A^2 -
B^2} B$. \\

As is well-known, Eqs. (\ref{Uomega},\ref{mnutbm}) lead to an ${\cal M}_{\nu f}=U(\omega)^T{\cal
M}_{\nu}U(\omega) $ in the form
exhibiting the TBM mixing:
\be \label{tnbmf}
{\cal M}_{\nu f}=\frac{1}{3a}
\left(
\begin{array}{ccc}
  (a+b)(3 a-b) & -b (a+b) & -b (a+b) \\
 -b (a+b) & b(2 a-b) & 3a^2-b(a+b) \\
 -b (a+b) & 3a^2-b(a+b) & b(2 a-b)
\end{array}
\right).
\ee

We need to change the existing model in two ways to obtain more predictive form of Eq.
(\ref{mnuf2}). First, we require that all the Yukawa couplings in Eq. (\ref{yukawa}) as well as the
vacuum expectation values are real. Eq. (\ref{tnbmf}) then coincides with the real part of
(\ref{mnuf2}) with 
\be \label{prmreal}
z=a-\frac{b}{3a}(a+b);~y=\frac{b}{3a}(2a-b);~x=-\frac{b}{3a}(a+b).\ee
We need to enlarge the model to introduce the imaginary part. This can be done either by adding
additional $SU(2)_L$ singlet or triplet fields transforming as an $A_4$ triplet. Conventionally, one
uses CP symmetry to obtain the real Yukawa couplings. The reality of Yukawa couplings follows if
definition of CP is generalized in a manner analogous to \cite{grimus1}. This generalized CP
combines the CP and $\mu$-$\tau$ symmetry as follows:
\beqa \label{cpdef}
(l_l,\nu_R,\Phi,\chi)&\rightarrow& S_3 (l_L^c,\nu_R^c,\Phi^\dagger,\chi^\dagger)~,\nonumber\\
(e_R,\mu_R,\tau_R)&\rightarrow& (e_R^c,\mu_R^c,\tau_R^c)~,\eeqa
where superfix $c$ on a field denotes its CP conjugate and  $S_3$ is defined in Eq. (\ref{s2s3}).
Note that the above symmetry behaves like ordinary CP on the $A_4$ singlet right handed charged
leptons and is thus slightly different from the generalized $\mu$-$\tau$ symmetry introduced in
\cite{trimaximal1}. 

The required imaginary part in ${\cal M}_{\nu f}$ can be generated  in two ways:
\subsection{Type-II extension}
Add three copies of $SU(2)_L$ triplet fields $\Delta$
which form a triplet of $A_4$ with the $U(1)_X$ charge  $X=2$. This modifies the Yukawa
interaction by an additional triplet seesaw term 
\be \label{yukawatriplet}
-{\cal L}_{Y}^{\Delta}= y_{L} (\overline{l}_L l_L^c)_{\bf 3} \Delta + h.c. \ee
$y_L$ becomes real if $\Delta\rightarrow  S_3\Delta^\dagger $ under the generalized CP.  
Let us now assume that $\Delta$ takes vev along the following direction
\be \label{deltavev}
\langle \Delta \rangle = \upsilon_{\Delta} (0,-1,1)^T
\ee
Such vacuum alignment can be achieved through some terms
that breaks $A_4$ softly and explicit example is discussed in \cite{maclassic}.
Eq. (\ref{yukawatriplet}) gives rise to a type-II contribution in neutrino
mass matrix. Combining this with the type-I contribution, Eq. (\ref{mnutbm}), we get the following:
\be \label{mnufnl1}
{\cal M}_{\nu}= \left(
\ba{ccc}
\frac{(a^2-b^2)}{a}&c&-c\\
c&a & b\\
-c&b & a\\
\ea
\right)
\ee
Parameters $a,b,c$  are real but when transformed to the flavour basis one obtains a complex
${\cal M}_{\nu f}$ coinciding exactly with Eq. (\ref{mnuf2}) with $x,y,z$ defined in Eq.
(\ref{prmreal}) and 
$$x'=-\frac{c}{\sqrt{3}}~.$$ 

The generalized $Z_2\times Z_2$ symmetry  emerges here as an 
effective symmetry. The type-II contribution (characterized by the parameter $c$) in the above neutrino mass matrix
generates nonzero reactor angle and modifies the solar mixing angle as in Eq. (\ref{solar2}). 

\subsection{Type-I extension}
Another viable extension of the model is obtained  by adding $A_4$ triplet, $SU(2)_L$ singlet field $\chi'$
in addition to $\chi$ already present. $\chi'$ introduces the following term in Eq. (\ref{yukawa}).
\be \label{yukawaflavon}
-{\cal L}_{Y}^{\chi'}= \frac{1}{2} y_{R} (\overline{\nu}_R \nu_R^c)_{\bf 3} \chi' + h.c. \ee
$y_{R}$ can be made real using the similar generalized CP symmetry defined in Eq. (\ref{cpdef}).
Assuming that $\chi'$ takes a vev along the same direction as $\Delta$ in the previous case,
{\it i.e.} $ \langle \chi' \rangle = \upsilon_{\chi'} (0,-1,1)^T $, we get 
\be \label{MR1}
M_R=\left(
\ba{ccc}
A&C&-C\\
C&A & B\\
-C&B & A\\
\ea
\right),
\ee
where $C=y_R \upsilon_{\chi'}$. After the seesaw the light neutrino mass matrix can be suitably
written as 
\be \label{mnufnl2}
{\cal M}_{\nu}= \left(
\ba{ccc}
\frac{(a^2-b^2+c^2)}{a}&c&-c\\
c&a & b\\
-c&b & a\\
\ea
\right).
\ee
This matrix also exhibits the generalized $Z_2\times Z_2$ symmetry and is determined by three real parameters as before. 
The only difference from the previous case is a small contribution of the $\sim {\cal O}
(a\theta_{13}^2)$ in 11 entry in ${\cal M}_{\nu}$. As a result the phenomenology of neutrino
masses in both cases are very similar and we now turn to this discussion.
\subsection{Phenomenology}
\begin{figure}[ht]
 \centering
\hspace*{-0.5cm}
\includegraphics[width=17.5cm]{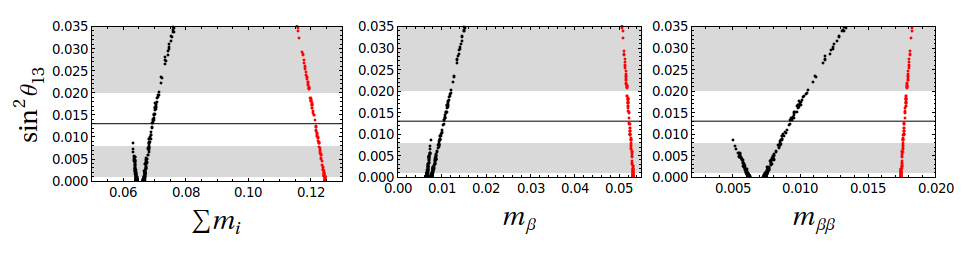}
\caption{Correlations between the reactor angle and different neutrino mass dependent observables implied by the 
neutrino mass matrix in Eq. (\ref{mnufnl1}). The black (red) points
correspond to the normal (inverted) hierarchy in neutrino masses. The black horizontal line shows
the mean value of $\sin^2\theta_{13}$ obtained from the global fits. The unshaded and the shaded
regions correspond to $1\sigma$ and $3\sigma$ ranges of $\sin^2\theta_{13}$ respectively.}
\label{fig1}
\end{figure}

We now derive the phenomenological consequences of the generalized $Z_2\times Z_2$ structures
Eqs. (\ref{mnufnl1},\ref{mnufnl2}) obtained using the $A_4$ symmetry and specific vacuum alignment.
While the most general, $Z_2\times Z_2$ invariant structure, Eq. (\ref{mnuf2}) has four parameters,
specific realization obtained here has only three parameters.  This follows from Eq. (\ref{prmreal})
which shows that   $x, y, z$ are not independent but are related by :
\be
z=-y+\frac{x(x+5y)}{2 x+y}~.\ee
The situation here is similar to the original $A_4$ models in which specific realizations of the TBM schemes
lead to more constrained mass pattern  than the most general one and lead to sum rules among
neutrino masses \cite{werner}. 
Specifically, Eq. (\ref{prmreal}) leads to a mass sum rule \cite{babuhe,werner}
\be \label{sumrule1}
\frac{2}{m_2}+\frac{1}{m_3}=\frac{1}{m_1}, \ee
where $m_i$ are the neutrino masses. Note that $m_i$ are real in our case since all the parameters
in the neutrino mass matrix (\ref{mnutbm}) are real. The phenomenological implications of this
neutrino mass sum-rule are already considered in \cite{babuhe,werner}. 
Generalization introduced through Eq. (\ref{mnufnl1}) modify this sum rule to
\be \label{sumrule2}
\frac{2}{m_2+3 (m_3-m_2)s_{13}^2 }+\frac{1}{m_3+3 (m_2-m_3) s_{13}^2}=\frac{1}{m_1}~. \ee
The above sum rule allows determination of the absolute neutrino mass scale as a function of $s_{13}^2$. This determination
depends on the type of hierarchy and approximate analytic form for the  lightest neutrino mass are given in the limit $s_{13}=0$
by \cite{werner}
\beqa \label{lightestnu}
{\text{For~normal~hierarchy}}~~~~~|m_1|&\approx&\sqrt{\frac{\dms}{3}}\left( 1 \pm
\frac{4\sqrt{3}}{9} \sqrt{\frac{\dms}{\Dma}}\right),\\
{\text{For~inverted~hierarchy}}~~~~|m_3|&\approx&\sqrt{\frac{\Dma}{8}}\left( 1 + \frac{1}{3}
\frac{\dms}{\Dma}\right).  \eeqa
Using the values of $\dms$ and $\Dma$ obtained from recent global fits \cite{valle} to the neutrino
oscillation data, above equations imply
\beqa \label{lightnumass}
{\text{For~normal~hierarchy}}~~~~~|m_1|&\approx& 5.7 \times 10^{-3}~{\rm eV}
~~~{\text{or}}~ \nonumber \\ 
 &\approx& 4.4 \times 10^{-3}~{\rm eV}, \\
{\text{For~inverted~hierarchy}}~~~~|m_3|&\approx&0.0179 ~{\rm eV}.  \eeqa
Further, the three mass dependent neutrino observables, namely (1) the sum of absolute neutrino
masses $\Sigma m_i$, (2) the kinematic electron neutrino mass in beta decay $m_\beta$ and (3) the
effective mass for the neutrinoless double beta decay $m_{\beta\beta}$ can also be obtained by their
approximated expressions given in \cite{werner,babuhe}. The presence of a non-zero $\theta_{13}$ modifies the
predicted values of observable compared to the models in \cite{werner,babuhe}. We  determine the effect of
non-zero $\theta_{13}$ numerically using  Eq. (\ref{mnufnl1}). The results of such analysis are
given in Fig. \ref{fig1}.\\

\begin{figure}[ht]
 \centering
\hspace*{-0.5cm}
\includegraphics[width=17.5cm]{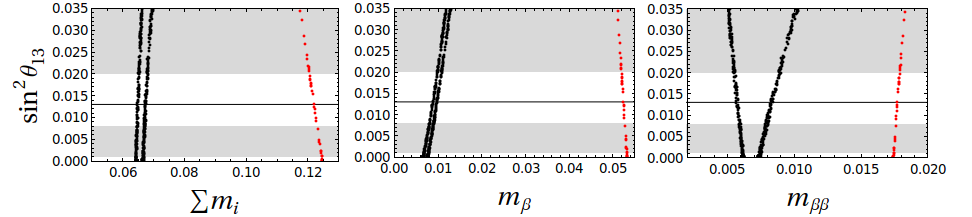}
\caption{Correlations between the reactor angle and different neutrino mass dependent observables
arises in the neutrino mass matrix given by Eq. (\ref{mnufnl2}). The black (red) points
correspond to the normal (inverted) hierarchy in neutrino masses. The black horizontal line shows
the mean value of $\sin^2\theta_{13}$ obtained from the global fits. The unshaded and the shaded
regions correspond to $1\sigma$ and $3\sigma$ ranges of $\sin^2\theta_{13}$ respectively.}
\label{fig2}
\end{figure}

As can be seen from Fig. \ref{fig1}, all the mass dependent observables varies slightly with
the reactor angle. These variations are smaller for inverted hierarchy compared to the normal
hierarchy. Results of a similar numerical analysis for purely type-I extension, Eq. (\ref{mnufnl2})
are given in Fig. \ref{fig2}.

\section{Summary}
\label{sec:summary}
The evidence of possible non-zero $\theta_{13}$ requires the modification of the TBM  patterns motivated by 
$A_4$ and other discrete symmetries. We have proposed a minimal modification which retains the prediction of the maximality
of $\theta_{23}$, allows non-zero $\theta_{13}$  and introduces small ${\cal O} (\theta_{13}^2)$ deviation from the 
$\theta_{12}$ predicted in the TBM. The 
basis of our proposal is the $Z_2\times Z_2$ symmetry of the TBM mass matrix. It is shown that combination of this symmetry 
with the CP gives rise to very predictive structure determined in terms of only four real parameters. The generalized 
$Z_2\times Z_2$ can emerge from a simple  extension of the standard $A_4$ schemes if Yukawa couplings and vev are real. The 
resulting neutrino mass matrix is quite predictive and is determined in terms of only three parameters making it one of the simplest
modification of the TBM scheme consistent with the present information on neutrino masses and mixing angles.

\noindent{\bf Acknowledgements}\\
ASJ thanks the Department of Science and Technology, Government of India for support under the J. C.
Bose National Fellowship programme, grant no. SR/S2/JCB-31/2010.

\end{document}